\documentclass[english,10pt]{article}

\usepackage[left=.81in,right=.81in,bottom=.85in,top=.80in]{geometry}

\usepackage[utf8]{inputenc}
\usepackage{babel}
\usepackage{slantsc}
\usepackage{array}
\usepackage{amsmath}
\usepackage{amsthm}
\usepackage{amssymb}
\usepackage{graphicx}
\usepackage{enumerate}
\usepackage{mathtools}
\usepackage{natbib}
\usepackage{bbm}
\usepackage{tikz}
\usepackage{multirow}
\usetikzlibrary{tikzmark}
\usepackage{xspace}
\usepackage{bigints}
\usepackage{booktabs}
\usepackage{comment}

\usepackage{pgf}
\usepackage{pgfplots}
\pgfplotsset{compat=newest} 
\pgfplotsset{plot coordinates/math parser=false} 
\newlength\figureheight 
\newlength\figurewidth 

\usepackage{heuristica}
\usepackage[heuristica,vvarbb,bigdelims]{newtxmath}
\usepackage[T1]{fontenc}

\newtheorem{theorem}{Theorem}
\newtheorem{proposition}{Proposition}

\newtheorem{corollary}{Corollary}

\newtheorem{lemma}{Lemma}

\newtheorem*{theorem*}{Theorem}
\theoremstyle{definition}
\newtheorem{definition}{Definition}
\newtheorem{example}{Example}
\newtheorem{assumption}{Assumption}

\newtheorem{remark}{Remark}

  \newtheorem*{rep@theorem}{\rep@title}
\newcommand{\newreptheorem}[2]{%
\newenvironment{rep#1}[1]{%
 \def\rep@title{#2 \ref{##1}}%
 \begin{rep@theorem}}%
 {\end{rep@theorem}}}
\newreptheorem{lemma}{Lemma}

\usepackage[labelsep=period]{caption}
\usepackage{subcaption}

\makeatletter
\newcommand{\subalign}[1]{%
  \vcenter{%
    \Let@ \restore@math@cr \default@tag
    \baselineskip\fontdimen10 \scriptfont\tw@
    \advance\baselineskip\fontdimen12 \scriptfont\tw@
    \lineskip\thr@@\fontdimen8 \scriptfont\thr@@
    \lineskiplimit\lineskip
    \ialign{\hfil$\m@th\scriptstyle##$&$\m@th\scriptstyle{}##$\crcr
      #1\crcr
    }%
  }
}
\makeatother

\definecolor{PennBlue}{RGB}{001,031,091}
\definecolor{PennRed}{RGB}{153,0,0}
\usepackage{xurl}
\usepackage{hyperref}
\hypersetup{
	pdfborder = {0 0 0},
    colorlinks,
    citecolor=PennRed,
    filecolor=PennRed,
    linkcolor=PennRed,
    urlcolor=PennRed
}

\AtBeginDocument{}

\author{A. Arda Gitmez\thanks{Department of Economics, Bilkent University. E-mail: \href{mailto:arda.gitmez@bilkent.edu.tr}{arda.gitmez@bilkent.edu.tr}.} \and Pooya Molavi\thanks{Northwestern University. E-mail: \href{mailto:pmolavi@kellogg.northwestern.edu}{pmolavi@kellogg.northwestern.edu}.} }
\title{Informational Autocrats, Diverse Societies\thanks{We would like to thank Daron Acemoglu, Raphael Boleslavsky, Georgy Egorov, Emin Karag\"{o}zo\v{g}lu, Nicola Persico, Mehdi Shadmehr, Emilie Sartre, Alex Wolitzky and Leifu Zhang for helpful discussions, and various seminar participants for their comments. Earlier versions of this paper were circulated under the title ``Media Capture: A Bayesian Persuasion Approach'' and ``Polarization and Media Bias.''
}
}
\date{\today}

\AtBeginDocument{%
  \mathchardef\mathcomma\mathcode`\,
  \mathcode`\,="8000 
}
{\catcode`,=\active
  \gdef,{\mathcomma\discretionary{}{}{}}
}
\vfuzz \hfuzz
\allowdisplaybreaks

 \hypersetup{
           breaklinks=true,   
           colorlinks=true,   
           pdfusetitle=true,  
        }

\begin{document}
\maketitle
\begin{abstract}
\fontsize{10}{10}\selectfont \baselineskip0.55cm
\noindent This paper presents a theoretical model of an autocrat who controls the media in an attempt to persuade society of his competence. We base our analysis on a Bayesian persuasion framework in which citizens have heterogeneous preferences and beliefs about the autocrat. We characterize the autocrat's information manipulation strategy when society is monolithic and when it is divided. When the preferences and beliefs in society are more diverse, the autocrat engages in less information manipulation. Our findings thus suggest that the diversity of attitudes and opinions can act as a bulwark against information manipulation by hostile actors.
\end{abstract}

\thispagestyle{empty}
\newpage 
\setcounter{page}{1}\fontsize{11.5}{11.5}\selectfont\baselineskip0.65cm

\section{Introduction} \label{section:introduction}
Over the past two decades, many democracies have devolved into hybrid regimes and outright autocracies \citep{bermeo:2016,levitsky/ziblatt:2018,haggard/kaufman:2021}. From Venezuela's Hugo Ch\'{a}vez to Hungary's Victor Orb\'{a}n and Russia's Vladimir Putin, politicians who came to power through democratic means have consolidated their control and undermined democratic institutions. Unlike the dictators of the 20th century, this new breed of autocrat does not resort to overt violence. Instead, they maintain power by building support among the masses and winning elections that appear to be democratic. To cultivate their image as competent leaders, they manipulate information by controlling state media \citep{rozenas/stukal:2019}, co-opting or pressuring independent media outlets \citep{mcmillan/zoido:2004,szeidl/szucs:2021}, and covertly censoring unfavorable news \citep{lorentzen:2014}.\footnote{Attempts to control media and manipulate information extend beyond autocracies. Examples in democracies abound, from Argentina \citep{ditella/franceschelli:2011} to Italy \citep{durante/knight:2012}, and from Mexico \citep{stanig:2015} to the United States \citep{qian/yanagizawadrott:2017,gentzkow/shapiro:2008}.} They are, as \cite{guriev/treisman:2019,guriev/treisman:2020} put it, {\it informational autocrats}.

But not all autocracies are alike. Even setting aside those that adhere to the 20th-century playbook and completely control the media (such as North Korea), there is still a wide variation in media freedom across informational autocracies. As \cite{egorov/sonin:2022} note, ``media freedom varies a lot across nondemocratic regimes, from levels comparable to mature democracies, to that of totalitarian regimes.''\footnote{A close look at the Global Media Freedom Dataset (GMFD) of \cite{whitten-woodring/vanbelle:2017} reveals the extent of variation. Among the 196 countries from 1948 to 2010 included in GMFD, 5275 country-year pairs are labeled as ``non-democracies'' based on their Polity Score. Interestingly, 4456 of such observations are classified as having ``Not Free Media'' by the authors, whereas the remaining observations have ``Imperfectly Free/Free Media'' \citep[Fig.1, p.184]{whitten-woodring/vanbelle:2017}.} The extent of information manipulation within a country also varies substantially over time. For instance, \cite{tai:2014} shows that from 2007 to 2013, the Chinese ``propaganda apparatus has banned fewer reports and guided more of them.'' As \citet{guriev/treisman:2019} observe, ``in Taiwan, an overt dictatorship under Chiang Kai-Shek evolved into an informational autocracy under his son, Chiang Ching-Kuo, in his later years, before transitioning to full democracy in the 1990s.''\footnote{Yet another example is the immense pressure on media outlets during Rafael Correa's rule in Ecuador from 2007 to 2017, and the backtracking of these policies by his vice-president and successor, Len\'{i}n Moreno: \url{https://www.cjr.org/analysis/ecuador-moreno-correa-supercom-press-freedom.php}.} This raises the question of why some societies are capable of preserving a degree of media freedom under autocratic rule while others are totally dominated by information manipulation.

This paper establishes a theoretical link between the distribution of attitudes and opinions in a society and its vulnerability to information manipulation. It argues that autocrats engage in less information manipulation in more diverse societies. Informational autocrats need to fine-tune their manipulation strategies to citizens' attitudes and opinions; greater diversity complicates this task. As attitudes and opinions become more dispersed, it becomes harder for autocrats to convince their opponents without alienating their supporters. They respond optimally by manipulating information less and allowing for a more free media landscape.

We present this insight using a Bayesian persuasion model, \`{a} la \citet*{kamenica/gentzkow:2011}, with a population of heterogeneous citizens. The autocrat commits to a public communication policy. Citizens observe the message drawn according to the policy and decide whether to support the autocrat or not. The extent to which the autocrat manipulates information depends on the distribution of citizens' attitudes and opinions. We analyze the autocrat's decision and characterize the optimal information manipulation policy. 

At the model's core is the trade-off faced by an informational autocrat. To gain support, he must convince citizens of his competence through information manipulation. He would do so by sending the message that things are ``good'' as frequently as possible. However, citizens understand that information is manipulated and only act based on the autocrat's communication when they find it informative. As the autocrat manipulates information more, fewer people act based on the autocrat's messages but those who do lend him a higher support. The optimal policy balances these two effects. 

The paper's main result establishes that when citizens' attitudes and opinions are more dispersed, the autocrat finds it optimal to engage in less information manipulation. The intuition is best illustrated by introducing some heterogeneity to a Bayesian persuasion model where citizens share identical attitudes and opinions. In the homogeneous model, the optimal strategy involves sending the ``bad'' message just frequently enough to make citizens indifferent between supporting the autocrat and not upon receiving the ``good'' message. However, in a society with dispersed attitudes and opinions centered around those in the homogeneous society, this strategy would only secure the support of half the citizens when the ``good'' message is sent. To gain broader support, the autocrat must appeal to more skeptical citizens, which requires a reduction in information manipulation. 

Our main contribution is to illustrate how this reasoning generalizes under a novel partial order on distributions, which captures the idea of dispersion in attitudes and opinions. We start with a model in which citizens are heterogeneous along two dimensions: preferences and initial beliefs about the autocrat. We collapse these two dimensions of heterogeneity into a one-dimensional distribution, which we call the {\it virtual density}. The virtual density is a sufficient statistic for the distribution of opinions and attitudes when it comes to citizens' support for the autocrat. When the virtual density is single-peaked, the opinions and attitudes in the society are similar to each other, i.e., the society is {\it monolithic}. Conversely, when the virtual density is single-dipped, there are two large groups with opposing attitudes toward the autocrat, i.e., the society is {\it divided}. We characterize the optimal persuasion policy in the cases where the virtual density is single-peaked and single-dipped. This allows us to consistently define what it means for the autocrat to engage in {\it more/less information manipulation}. It also allows us to introduce a partial order on one-dimensional distributions that captures the idea of {\it dispersion}. Our comparative statics results establish that when the virtual density is more dispersed, the autocrat engages in less information manipulation under the optimal policy. This result holds in both monolithic and divided societies.

\paragraph*{Related Literature.} First and foremost, our model contributes to the growing literature on informational autocrats \citep*{guriev/treisman:2019,guriev/treisman:2020,guriev/treisman:2022,egorov/sonin:2022,gehlbach/luo/shirikov/vorobyev:2022}.\footnote{Also related is the literature on {\it democratic authoritarianism} \citep{brancati:2014} and {\it competitive authoritarianism} \citep{levitsky/way:2002} in political science. However, those works are less focused on information manipulation and more on the dismantling of democratic institutions.} A closely related literature studies {\it media capture}, the idea that politicians exert control over media by co-opting private media \citep{besley/prat:2006}, controlling state media \citep{gehlbach/sonin:2014}, censoring news \citep*{shadmehr/bernhardt:2015,boleslavsky/shadmehr/sonin:2021}, or controlling media's access to information \citep{ozerturk:2022}; see \cite{prat:2015}  and \cite{enikolopov/petrova:2015} for two comprehensive reviews.\footnote{\cite{corneo:2006, petrova:2008, petrova:2012}, and \cite{alonso/padro-i-miquel:2022} present models of media capture by special interest groups.} We contribute to this literature by establishing that the vulnerability of a society to media capture depends not only on the opinion of the median citizen but also on the dispersion of opinions in society.

Another literature focuses on understanding the variation in information manipulation. As a source of variation, \cite*{egorov/guriev/sonin:2009} study the natural resource endowment, \cite{vondoepp/young:2013} study the threats that governments face, while \cite{mcgreevy-stafford:2020} study protests. A closely related literature identifies factors that limit information manipulation. \cite*{ditella/galiani/schargrodsky:2012} emphasize first-hand experiences, \cite{durante/knight:2012}, \cite{glassel/paula:2020}, \cite{knight/tribin:2022}, and \cite*{enikolopov/rochlitz/schoors/zakharov:2023} emphasize the existence of alternative media outlets, \cite*{qin/stromberg/wu:2018} emphasize market competition, and \cite{knight/tribin:2019} emphasize citizens' ability to ``tune out.'' Our findings contribute to this literature by highlighting the role of diversity of citizens' attitudes and opinions in limiting the extent of information manipulation.

We also contribute to the literature on political consequences of diversity in a society. Unlike the literature on the impact of heterogeneity on conflict \citep*{desmet/ortuno-ortin/wacziarg:2017,arbatli/ashraf/galor/klemp:2020} and political institutions and governance \citep*{alesina/baqir/easterly:1999,laporta/lopez-de-silanes/shleifer/vishny:1999,lindqvist/ostling:2010,galor/klemp:2018}, our focus is on the impact of population heterogeneity on information manipulation.

Our model of information manipulation follows the Bayesian persuasion approach \citep*{kamenica/gentzkow:2011}. We contribute to this literature by allowing for receivers with both heterogeneous preferences \citep*{wang:2015,alonso/camara:2016aer,kolotilin/li/myloyanov/zapechelnyuk:2017,bardhi/guo:2018,chan/gupta/li/wang:2019,arieli/babichenko:2019,kerman/herings/karos:2021,sun/schram/sloof:2022} and heterogeneous priors \citep*{alonso/camara:2016jet, laclau/renou:2017,kosterina:2022,innocenti:2022,gitmez/sonin:2023}.\footnote{Also related is the literature on information design, which studies the optimal information structure in a game to be played among multiple players \citep*{bergemann/morris:2019,taneva:2019,mathevet/perego/taneva:2020,inostroza/pavan:2022}.} To collapse our two-dimensional primitives onto a single dimension, we introduce an object that we call virtual density. Our comparative statics results rely on the partial order we introduce on the virtual density. \citet*{kolotilin:2015}, \citet*{kolotilin/mylovanov/zapechhelnyuk:2022}, \citet*{sun/schram/sloof:2022}, and \citet*{curello/sinander:2022} also conduct comparative statics exercises in Bayesian persuasion settings. Whereas \citet*{kolotilin:2015} focuses on changes in welfare, we analyze how the optimal policy changes with parameters of the model. \citet*{sun/schram/sloof:2022} derive comparative statics results with respect to the sender's preferences. Our comparative statics result complement theirs by focusing on changes in the receivers' characteristics. \citet*{kolotilin/mylovanov/zapechhelnyuk:2022}'s comparative statics are with respect to the receivers' inclination to be persuaded, whereas ours are with respect to the receivers' heterogeneity. Finally, in parallel work, \citet*{curello/sinander:2022} examine the comparative statics of Bayesian persuasion. However, our approach differs from theirs in two key ways. Firstly, while they focus on changes in the sender's value function, we focus on variations in the distribution of receiver types. Secondly, our approach employs a dispersion partial order on distributions, enabling us to view changes in receiver types as changes in the extent of their heterogeneity.

A crucial assumption in the Bayesian persuasion literature is  commitment by the sender. In our model, the sender can commit to a public communication policy, which is observed by all receivers. This assumption can be defended on several grounds. First of all, in our setup, persuasion satisfies the credibility assumption of \citet*{lin/liu:2021}.\footnote{In particular, we can allow for undetectable deviations by the sender. Since the sender's payoff in our model is additively separable, there is no profitable deviation that gives the same message distribution as the optimal policy. It should be noted that, with heterogeneous priors, the set of undetectable deviations is different for each agent. In such a case, one has to require deviations to be undetectable given the sender's prior.} Second, the autocrat's policy can viewed as an ``editorial policy,'' which describes the general attitude of media sources, with the details of the coverage to be decided by reporters and editors \citep*{gehlbach/sonin:2014}. Finally, the outcome under commitment can be seen as a benchmark that describes the best-case scenario for the sender. Under this interpretation, our results characterize an ``ideal media landscape'' for a politician in a diverse society.

\section{Setup} \label{section:setup}

\subsection{The Model}
There are two types of agents: an autocrat and a unit measure of citizens, indexed by $r\in[0,1]$. Each citizen takes an action $a_r\in\{0,1\}$, the action being whether she supports the autocrat.

There is an underlying state of the world: $\theta \in \{0,1\}$. We call the $\theta = 1$ state the ``good'' state. When the state is good, supporting the autocrat is in  the citizens' best interest. The $\theta = 0$ state is the ``bad'' state where it is optimal for citizens not to support the autocrat. Citizen $r$'s payoff when she chooses action $a_r$ and the state is $\theta$ is given by
\begin{align} \label{eqn:utilitycitizen}
u_r(a_r,\theta) = a_r (\theta-c_r),
\end{align}
where $c_r \in [0,1]$ is citizen $r$'s cost of supporting the autocrat. If citizen $r$ knew the state, she would support the autocrat (i.e., $a_r=1$) in the good state and not support her (i.e., $a_r=0$) in the bad state. Hence, the good state may be interpreted as the state where the autocrat is competent and the bad state as the one where the autocrat is incompetent.

The autocrat wants to persuade citizens to support him regardless of the state of the world. 
His payoff when citizen $r$ provides support $a_r$ and the state is $\theta$ is given by
\begin{align} \label{eqn:utilityautocrat}
u_s(\{a_r\}_{r}) = \int_{0}^1 a_r dr.
\end{align}
When the state is good, the autocrat and citizens have common interests, whereas when the state is bad, their interests are opposed. We denote the autocrat's prior that the state is good  by $p_s\in(0,1)$.

Citizens do not learn the state of the world until after they have decided on whether to support the autocrat. Since citizens do not observe the state, they can only act based on their beliefs. The autocrat can influence those beliefs (and the resulting actions) by sending informative messages. To simplify the analysis, we assume that the autocrat can commit to a public communication strategy $\sigma: \{0,1\} \rightarrow \Delta(M)$, where $\sigma(\theta)[m]$ is the probability that public message $m\in M$ is generated when the state is $\theta$. The communication strategy represents the policies followed by media controlled by the autocrat and used by him to influence the views of citizens.

Citizens are heterogeneous in both their preferences and their prior beliefs. The heterogeneity of priors captures the idea that even people with identical payoffs may have different perspectives about the likelihood that the autocrat is competent. We let $p_r$ denote citizen $r$'s prior that the state is good and, and let $f(c,p)$ denote the joint density of costs and priors in the population. We take $f$ as a primitive of the model and study how changing this distribution affects the autocrat's optimal policy.  We assume that $f$ is common knowledge and continuously differentiable and bounded over its support.

The heterogeneity of perspectives poses a challenge for an autocrat who wants to garner broad support. Convincing different citizens with different preferences and beliefs requires different communication strategies. Yet, communication is public, so the autocrat cannot tailor his messaging strategy to citizens' diverse perspectives. The main characterization result of the paper concerns the optimal way of resolving the inherent tension in convincing different segments of the population.

\paragraph{Timing.} The timing of the communication game is as follows:
\begin{enumerate}
\item The prior and cost of each citizen are drawn, and citizen $r$ observes $(p_r,c_r)$.
\item The autocrat commits to a strategy $\sigma$, which is observed by all citizens.
\item The state is realized, and the autocrat sends the message drawn according to $\sigma$.
\item Each citizen $r$ updates her prior and chooses an action $a_r$.
\item Payoffs are realized.
\end{enumerate}
The solution concept we adopt is the Perfect Bayesian Equilibrium.

\subsection{An Equivalent Representative-Citizen Problem}
The fact that the autocrat is communicating with a population of heterogeneous citizens complicates her problem. However, the autocrat's optimal strategy can be found by solving a related persuasion problem with a \emph{representative citizen} whose prior coincides with the autocrat's prior. 

The key simplification comes from Proposition 1 of \citet*{alonso/camara:2016jet}. Consider citizens $r$ and $r'$ with priors $p_r$ and $p_{r'}=p_s$. Since the two citizens observe the same (public) message, their posteriors are related through the following expression:
\begin{align} \label{eqn:posteriormapping}
    \mu_r = \frac{\mu_{r'} \frac{p_r}{p_{r'}}}{\mu_{r'} \frac{p_r}{p_{r'}} + (1-\mu_{r'})\frac{1-p_r}{1-p_{r'}}},
\end{align}
where $\mu_r$ and $\mu_{r'}$ denote the posteriors of $r$ and $r'$, respectively.\footnote{Throughout the paper, we use posterior to mean subjective posterior probability of state $\theta=1$ given an agent's information.} This coupling of posteriors holds regardless of the communication strategy employed by the autocrat. It uniquely pins down the posterior $\mu_r$ of \emph{every} citizen $r$ as a function of the posterior of citizen $r'$---who will be our representative citizen.

Citizen $r$ supports the autocrat if and only if her posterior that the state is good is at least as large as his cost of action; that is, $a_r = 1$ if and only if
\begin{align} \label{eqn:cbound}
  c_r \leq c(\mu_s,p_r) \equiv \frac{\mu_s\frac{p_r}{p_s}}{\mu_s\frac{p_r}{p_s} + (1-\mu_s)\frac{1-p_r}{1-p_s}},
\end{align}
where $\mu_s$ denotes the posterior of the representative citizen (who has the same prior as the autocrat). The payoff to the autocrat is the share of the population who supports him:
\begin{align} \label{eqn:valuefunction}
    v(\mu_s) = \int_{0}^{1} \int_{0}^{c(\mu_s,p)} f(p,c) dc  dp.
\end{align}

The autocrat's problem is thus equivalent to a standard Bayesian persuasion problem with a representative citizen. The autocrat and representative citizen share the common prior $p_s$ that the state is good. The payoff to the autocrat, when he induces a posterior of $\mu_s$ for the representative citizen, is given by $v(\mu_s)$, defined in equation \eqref{eqn:valuefunction}. Following \citet*{kamenica/gentzkow:2011}, we refer to $v(\mu_s)$ as the autocrat's \emph{value function}. Whenever there is no risk of confusion, we drop the $s$ subscript and simply write $v(\mu)$ for the value to the autocrat of inducing posterior $\mu$ for the representative citizen.

The value function has several useful properties. First, $v(\mu)$ is increasing in $\mu$. Inducing a higher posterior for the representative citizen results in a higher posterior for every citizen, thus increasing the share of citizens who support the autocrat. Second, $v(0)=0$ and $v(1)=1$. When the representative citizen is certain that the state is bad, so is every other citizen. Therefore, no citizen supports the autocrat. Likewise, when the representative citizen is certain that the state is good, every other citizen is also certain that the state is good and supports the autocrat. Finally, $v(\mu)$ is differentiable in $\mu$ due to the differentiability of $f$.

The value function can thus be seen as a differentiable cumulative distribution function. We let $h(\mu)\equiv v'(\mu)$ denote the corresponding density and refer to it as the \emph{virtual density} of the persuasion problem with heterogeneous citizens. The virtual density has an intuitive interpretation: $h(\mu)$ is the density of citizens who are indifferent between taking the two actions whenever the representative citizen's posterior is equal to $\mu$. Importantly, the construction of $h$ allows us to collapse a two-dimensional object into a one-dimensional one. It defines a {\it belief threshold} for every citizen such that citizen $r$ supports the autocrat if and only if the representative citizen's posterior is above citizen $r$'s threshold. Note that the virtual density is a primitive of the problem---the shape of $f$ determines the shape of $h$. For example, when citizens have a common prior that coincides with the autocrat's prior, the virtual density reduces to the probability density $f(c)$ of costs in the population.

\section{Information Manipulation in Monolithic Societies} \label{section:monolithic}

\subsection{Single-Peaked Distributions}

The solution to the autocrat's persuasion problem takes a particularly simple form when the distribution of citizen types satisfies the following condition:
\begin{definition}\label{ass:single-peakedness}The virtual density $h(\mu)$ is \emph{single-peaked} if there exists some $\tilde{\mu}\in[0,1]$ such that $h'(\mu)>0$ for all $\mu<\tilde{\mu}$ and $h'(\mu)<0$ for all $\mu> \tilde{\mu}$.
\end{definition}
Single-peakedness is an assumption on the joint distribution of citizens' costs and prior beliefs. It requires a large share of citizens to have moderate preferences and beliefs, with fewer and fewer people having extreme preferences or beliefs. We thus consider single-peaked virtual densities to be representative of {\it monolithic} societies.

The significance of Definition \ref{ass:single-peakedness} rests on the following observation: when the virtual density is single-peaked, the autocrat's value function is first convex and then concave. Therefore, Corollary 2 of \citet*{kamenica/gentzkow:2011} implies the following characterization of the optimal strategy:

\begin{proposition} \label{proposition:single-peaked}
If the virtual density is single-peaked, the optimal strategy uses only two messages, and one of the messages fully reveals the bad state.
\end{proposition}

We maintain the assumption of single-peakedness throughout this section. We do so in part for tractability. However, single-peaked distributions also constitute a natural and widely used class of distribution functions. In Section \ref{section:single-dipped}, we show that the optimal strategy in the case where the virtual density is instead single-dipped is the mirror image of the optimal strategy in the single-peaked case. 

Whether the virtual density is single-peaked only depends on the distribution of types, $f$, and the autocrat's prior, $p_s$. In the remainder of this subsection, we find a set of easy-to-check sufficient conditions for the virtual density to the single-peaked. If citizens have a common prior that coincides with the autocrat's prior, then the single-peakedness of the virtual density is equivalent to the single-peakedness of the density of costs:

\begin{proposition} \label{proposition:commonpriors-singlepeaked}
Suppose $p_r=p_s$ for all $r$. The virtual density $h(\mu)$ is single-peaked in $\mu$ if and only if the density of costs $f(c)$ is single-peaked in $c$.
\end{proposition}

If citizens instead have a common cost, then the single-peakedness of the virtual density is implied by a condition that is weaker than the log-concavity of the density of priors:

\begin{proposition}\label{proposition:commoncost-logconcave}
Suppose $c_r=c\in(0,1)$ for all $r$. The virtual density $h(\mu)$ is single-peaked if the density of priors $f(p)$ is strictly positive for all $p\in(0,1)$ and satisfies
\begin{align} \label{eqn:commoncost-logconcave}
    \frac{d^2}{d p^2} \log f(p) < 2 \left(\gamma-1\right)^2  \min\bigg\{1,\frac{1}{\gamma^2}\bigg\} \quad \text{for all }p\in(0,1),
\end{align}
where $\gamma \equiv \frac{1-c}{c} \frac{1-p_s}{p_s}\geq 0$.
\end{proposition}

The following corollary of Proposition \ref{proposition:commoncost-logconcave} is a straightforward consequence of the facts that the left-hand side of equation \eqref{eqn:commoncost-logconcave} is negative if $f(p)$ is log-concave, while its right-hand side is always non-negative:\footnote{See \citet*{bagnoli/bergstrom:2005} for a list of well-known distributions satisfying log-concavity.}
\begin{corollary}
Suppose $c_r=c$ for all $r$. The virtual density $h(\mu)$ is single-peaked in $\mu$ if the density of priors $f(p)$ is strictly log-concave in $p$.
\end{corollary}

\subsection{A Measure of Information Manipulation}

In light of Proposition \ref{proposition:single-peaked}, we can assume without loss that the autocrat uses only two messages. We label the messages $m \in M = \{0,1\}$, with $m = 1$ the ``good'' message, which is suggestive of $\theta =1$, and $m = 0$ the ``bad'' message, which is suggestive of $\theta =0$. The autocrat's strategy can be represented by a pair of numbers:
\begin{align*}
\sigma = (\sigma^0, \sigma^1) \in [0,1]^2,
\end{align*}
where $\sigma^\theta \equiv \sigma(\theta)[m=1]$ is the probability of sending the good message in state $\theta \in \{0,1\}$. Throughout, we assume without loss of generality that $\sigma^1 \geq \sigma^0$. 

The autocrat \emph{manipulates information} if he sends the good message when the state is bad or sends the bad message when the state is good. By Proposition \ref{proposition:single-peaked}, when the virtual density is single-peaked, the bad message fully reveals the bad state; this entails sending the good message whenever the state is good, i.e., $\sigma^1 = 1$. Therefore, in the single-peaked case, the extent of information manipulation is conveniently summarized by the probability $\sigma^0$ of sending the good message when the state is bad. We use the following notion of information manipulation in this case:
\begin{definition}
Consider single-peaked virtual densities $h_1$ and $h_2$ with the corresponding optimal strategies $\sigma_1=(\sigma^0_1,\sigma^1_1)$ and $\sigma_2=(\sigma^0_2,\sigma^1_2)$ for the autocrat. The autocrat \emph{manipulates information less} given $h_1$ than given $h_2$ if $\sigma^0_1\leq \sigma^0_2$.
\end{definition}

\subsection{A Measure of Dispersion} \label{section:polarization}
To study how diversity affects information manipulation, we need to introduce a measure of dispersion. Our measure is a novel partial order on probability distributions:
\begin{definition}
Consider two single-peaked distributions with densities $f_1$ and $f_2$ supported on a common compact set. $f_1$ is \emph{more dispersed} than $f_2$ if
\begin{equation}
f_2(x) = \alpha\left(f_1(x)\right) \qquad \text{for all}\; x\label{eqn:virtualdensity-polarization},
\end{equation}
for some strictly increasing and convex function $\alpha: \mathbb{R}^+ \rightarrow \mathbb{R}^+$. 
\end{definition} 

\begin{figure}[htbp!]
\centering\scalebox{.65}{\input{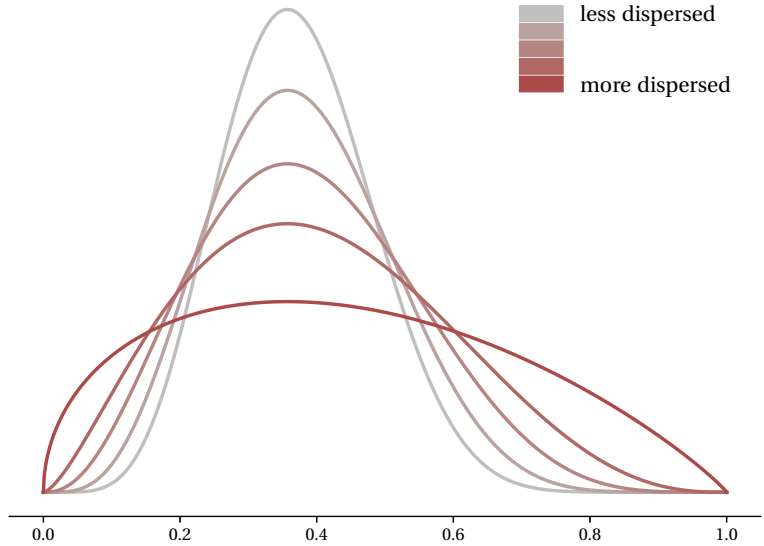}}\caption{The dispersion order on single-peaked densities.}\label{fig:polarization_sp}
\end{figure}

This partial order has an intuitive interpretation. Since $\alpha$ is increasing, $f_2$ is single-peaked whenever $f_1$ is single-peaked. Since $\alpha$ is convex and $f_1$ and $f_2$ both have to integrate to one, transforming $f_1$ by $\alpha$ magnifies the parts of $f_1$ with larger values and shrinks the parts with smaller values. Moving from $f_1$ to $f_2$ thus moves mass from parts of the distribution that initially have smaller mass to parts with larger initial mass. In other words, $f_2$ looks like $f_1$, but with higher peaks and deeper troughs. But since $f_1$ is single-peaked, most of its mass is concentrated around its peak. Therefore, $f_2$ has even more mass in the center and even less mass in the periphery relative to $f_1$; that is, $f_1$ is more dispersed than $f_2$. Figure \ref{fig:polarization_sp} illustrates the probability density functions for a set of single-peaked Beta distributions that are ranked in the dispersion order.

Members of many parametric families of distributions can be ordered in the dispersion order. Two examples follow:
\begin{example}
Consider two single-peaked Beta distributions:
\begin{align*}
    f_1 & = \text{Beta}(\alpha_1, \beta_1),\\
    f_2 & = \text{Beta}(\alpha_2, \beta_2),
\end{align*}
where $\frac{\alpha_1-1}{\alpha_1 + \beta_1 -2} = \frac{\alpha_2-1}{\alpha_2 + \beta_2 -2}$. If $\alpha_1 \geq \alpha_2$, then $f_2$ is more dispersed than $f_1$, while if $\alpha_1\leq \alpha_2$, then $f_1$ is more dispersed than $f_2$. In particular, any two single-peaked Beta distributions with the same mode are ranked according to the dispersion partial order.
\end{example}

\begin{example}
Consider the following truncated normal distributions on $[0,1]$:
\begin{align*}
    f_1 & = \text{TruncatedNormal}(\mu, \sigma^2_1),\\
    f_2 & = \text{TruncatedNormal}(\mu, \sigma^2_2).
\end{align*}
If $\sigma^2_2 \geq \sigma^2_1$, then $f_2$ is more dispersed than $f_1$.
\end{example}

\cite{johnson/myatt:2006}'s {\it rotation order} is a related partial order, which also ranks distributions in terms of their dispersion or heterogeneity. The main difference between the two is that \cite{johnson/myatt:2006} consider rotations of a cumulative distribution function around a given point, whereas in our partial order the rotation point itself depends on the distribution function. The endogeneity of the rotation point to the distribution function is crucial for our comparative statics results. It ensures that the rotation point is always in the appropriate range for an increase in dispersion to have an unambiguous effect on the extent of information manipulation.

\subsection{Dispersion and Information Manipulation in Monolithic Societies}

We are now ready to examine how dispersion affects information manipulation. Our main result establishes that information manipulation is less severe in more diverse societies:

\begin{theorem} \label{theorem:sc-virtualdensitypolarization}
Let $h_1$ and $h_2$ be two single-peaked virtual densities. If $h_1$ is more dispersed than $h_2$, then the autocrat manipulates information less given $h_1$ than $h_2$.
\end{theorem}

To gain some intuition, recall that each citizen is assigned a belief threshold, which ranks them by their inclination to support the autocrat---those with lower thresholds are more supportive. The autocrat's optimal strategy targets a marginal citizen who, upon receiving the positive message, is just indifferent between supporting the autocrat and not. Citizens with lower thresholds act in line with the message, while those with higher thresholds never support the autocrat. The optimal strategy balances the autocrat's goal of maximizing the mass of supporters with that of maximizing the support frequency. In a less dispersed society, the belief thresholds are tightly concentrated around the modal citizen's threshold. Therefore, targeting a citizen whose threshold is slightly above the mode ensures the support of almost every citizen. But in a diverse society with more dispersed thresholds, this same approach yields too few supporters. To counter this, the autocrat needs to increase the informativeness of the media, appealing to those with belief thresholds further from the mode.

Theorem \ref{theorem:sc-virtualdensitypolarization} describes the impact of dispersion on information manipulation while maintaining the assumption that the society is monolithic, and so, the virtual density is single-peaked. In the next section, we study persuasion in highly divided societies, in which there are more people in the extremes of preference and belief distribution than at its center.

\section{Divided Societies}\label{section:single-dipped}

Throughout this section, we study the properties of the optimal persuasion strategy when the virtual density is the polar opposite of single-peaked:

\begin{definition}\label{ass:single-dippedness}
The virtual density $h(\mu)$ is \emph{single-dipped} if there exists some $\tilde{\mu}\in[0,1]$ such that $h'(\mu)<0$ for all $\mu<\tilde{\mu}$ and $h'(\mu)>0$ for all $\mu>\tilde{\mu}$.
\end{definition}
In a society with a single-dipped virtual density, there are fewer moderates than those with extreme preferences or beliefs. Therefore, we consider single-dipped virtual densities to be representative of {\it divided} societies.\footnote{Following \citet[Figure 1]{fiorina/abrams:2008}, one may also call such a society {\it polarized}. We refrain from adopting this terminology, because polarization is typically visualized as the society having a small number of groups, with high homogeneity within groups and high heterogeneity across groups \citep{esteban/ray:1994}.}

As in the single-peaked case, the single-dippedness of the virtual density only depends on the distribution of types and the autocrat's prior. For example, if citizens have a common prior that coincides with the autocrat's prior, the single-dippedness of the virtual density is equivalent to the single-dippedness of the density of costs:
\begin{proposition}\label{proposition:commonpriors-singledipped}
Suppose $p_r=p_s$ for all $r$. The virtual density $h(\mu)$ is single-dipped in $\mu$ if and only if the density of costs $f(c)$ is single-dipped in $c$.
\end{proposition}

If citizens have a common cost, the single-dippedness of the virtual density is implied by a condition that is stronger than the log-convexity of the density of priors:

\begin{proposition}\label{prop:commoncost-logconvex}
Suppose $c_r=c$ for all $r$. The virtual density $h(\mu)$ is single-dipped if 
\begin{align} \label{eqn:commoncost-logconvex}
    \frac{\partial^2}{\partial p^2} \log f(p) > 2 \left(\gamma-1\right)^2  \max\bigg\{1,\frac{1}{\gamma^2}\bigg\} \quad \text{for all }p\in[0,1],
\end{align}
where $\gamma \equiv \frac{1-c}{c} \frac{1-p_s}{p_s}\geq 0$.
\end{proposition}
Note that the right-hand side of equation \eqref{eqn:commoncost-logconvex} is positive, and the left-hand side is positive if $f(p)$ is log-convex. Therefore, condition \eqref{eqn:commoncost-logconvex} can be interpreted as ``$f(p)$ being sufficiently log-convex.'' If $p_s + c = 1$, then $\gamma = 1$, and condition \eqref{eqn:commoncost-logconvex} reduces to the log-convexity of the distribution of priors.

When the virtual density is single-dipped, the autocrat's value function is first concave and then convex. Our next result characterizes the optimal persuasion strategy in this case.

\begin{proposition} \label{proposition:single-dipped}
If the virtual density is single-dipped, the optimal strategy uses only two messages, and one of the messages fully reveals the good state.
\end{proposition}

A comparison of Propositions \ref{proposition:single-peaked} and \ref{proposition:single-dipped} reveals that the optimal persuasion strategies are qualitatively different in monolithic and divided societies. In a divided society, there are many strong supporters (i.e., those with belief thresholds close to zero) and many strong skeptics (i.e., those with belief thresholds close to one). The autocrat's challenge is to convince the skeptics without alienating his supporters. The optimal strategy is to use a media source that frequently sends the bad message, so that the rare but credible occurrence of the good message is sufficient to convince even the most skeptical citizens.\footnote{One may interpret the optimal persuasion strategy as the existence of a limited number of independent media that are often critical of the autocrat. The generally critical coverage by such media lends them credibility, allowing the autocrat to benefit from their positive coverage in times of crisis. Such strategies are indeed employed by informational autocrats from time to time. For instance, following the anti-government protests and riots in Zhanaozen in December 2011, Kazakhstan's President Nursultan Nazarbayev suffered from a lack of credibility of the state broadcasting outlets. When all else failed to calm the public, the government invited six well-known bloggers, most labeling themselves as ``independent,'' to make a two-day visit to Zhanaozen. The bloggers carried a sense of credibility that the government sources lacked, and they were ``quite effective at reassuring readers that the city was outwardly calm, that rumors of morgues or hospitals full of corpses were unfounded and that shops were well-stocked and inhabitants able to buy food and drink'' \citep[p.267, also see \citealp{guriev/treisman:2022}, p.79]{lewis:2016}. In a similar episode, Vladimir Putin utilized the liberal Russian radio station Echo of Moscow to cover a credible account of a large pro-government demonstration in the capital in early 2012, thereby discouraging participation in opposition rallies elsewhere \citep{sobolev:2023}.}\footnote{\cite{baum/groeling:2009}, \cite{ladd/lenz:2009}, and \cite{chiang/knight:2011} document evidence of the persuasive power of communication when messages are sent by actors least expected to send them.} The strong supporters then have no reason to follow the media, because they support the autocrat even when the bad message is realized.

When the virtual density is single-dipped, the autocrat's optimal strategy entails sending the bad message whenever the state is bad. Then, the extent of information manipulation is summarized by the probability $\sigma^1$ of sending the good message when the state is good:
\begin{definition}
Consider single-dipped virtual densities $h_1$ and $h_2$ with the corresponding optimal strategies $\sigma_1=(\sigma^0_1,\sigma_1^1)$ and $\sigma_2=(\sigma^0_2,\sigma^1_2)$ for the autocrat. The autocrat \emph{manipulates information less} given $h_1$ than given $h_2$ if $\sigma_1^1\geq \sigma^1_2$.
\end{definition} 

We now examine the impact of increased dispersion on information manipulation. The following partial order is the appropriate adaptation of the partial order defined in Section \ref{section:polarization} for single-peaked densities to the set of single-dipped densities:
\begin{definition}

Consider two single-dipped distributions with densities $f_1$ and $f_2$ supported on a common compact set. $f_2$ is \emph{more dispersed} than $f_1$ if 
\begin{equation}
f_2(x) = \alpha\left(f_1(x)\right) \qquad \text{for all}\; x\label{eqn:virtualdensity-polarization_SD},
\end{equation}
for some strictly increasing and convex function $\alpha: \mathbb{R}^+ \rightarrow \mathbb{R}^+$.
\end{definition}

\begin{figure}[htbp!]
\centering\scalebox{.5}{\input{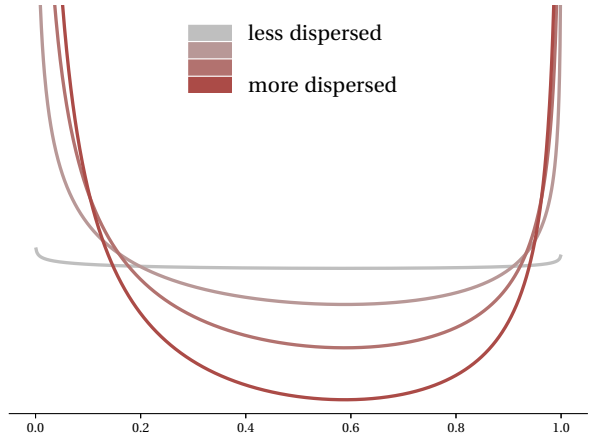}}\caption{The dispersion order on single-dipped densities.}\label{fig:polarization_sd}
\end{figure}

Figure \ref{fig:polarization_sd} illustrates the dispersion partial order on a set of a single-dipped Beta distributions. As the distribution becomes more dispersed, mass is moved from the center of the distribution to its tails.

Our next result establishes that, here, as in the single-peaked case, dispersion reduces information manipulation:

\begin{theorem} \label{theorem:sd-virtualdensitypolarization}
Let $h_1$ and $h_2$ be two single-dipped virtual densities. If $h_1$ is more dispersed than $h_2$, then the autocrat manipulates information less given $h_1$ than $h_2$.
\end{theorem}

Theorem \ref{theorem:sd-virtualdensitypolarization} shows that the main message of Theorem \ref{theorem:sc-virtualdensitypolarization} continues to hold in divided societies: Dispersion of attitudes and opinions reduces the extent of information manipulation. 

\begin{figure}[htbp!]
\centering\scalebox{.65}{\input{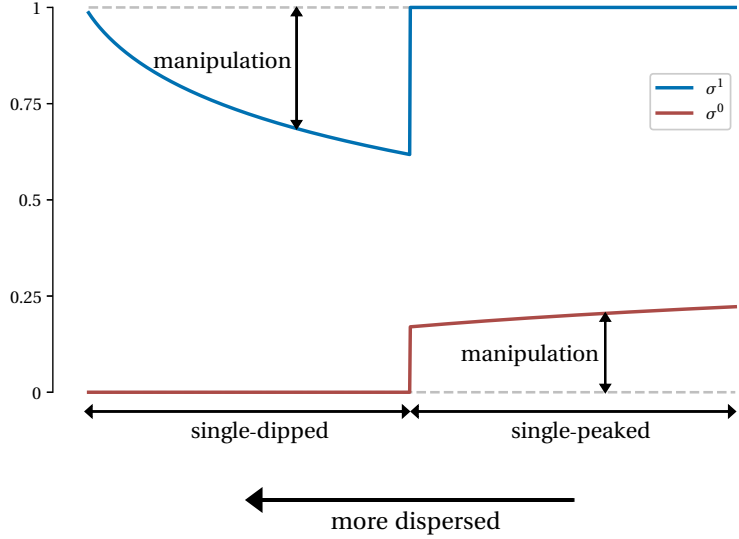}}\caption{Autocrat's information manipulation as a function of dispersion in society.}\label{fig:bias}
\end{figure}

The effect of dispersion on information manipulation can be succinctly summarized in a single figure by considering a parametric family of distributions that span both single-peaked and single-dipped cases. Figure \ref{fig:bias} illustrates the effect of dispersion on information manipulation for the case where the virtual density is a (symmetric) $\text{Beta}(1+\alpha,1+\alpha)$ distribution, and the autocrat's prior is given by $p_s=0.4$. The figure plots how the autocrat's optimal strategy changes as $\alpha$ ranges from $-1$ to $+1$. In the right half of the figure, $\alpha>0$, the distribution is single-peaked, and so, by Proposition \ref{proposition:single-peaked}, the optimal policy has the form $(\sigma_{\alpha}^0,\sigma_{\alpha}^1) = (\sigma_\alpha^0,1)$. As the society becomes more dispersed, by Theorem \ref{theorem:sc-virtualdensitypolarization}, $\sigma_{\alpha}^0$ decreases and the autocrat manipulates information less. On the left half of the figure, $\alpha<0$, the distribution is single-dipped, the optimal policy has the form $(\sigma_\alpha^0,\sigma_\alpha^1) = (0,\sigma_\alpha^1)$ (by Proposition \ref{proposition:single-dipped}), and $\sigma_\alpha^1$ increases and information manipulation decreases with dispersion (by Theorem \ref{theorem:sd-virtualdensitypolarization}).\footnote{Transitioning from a single-peaked to a single-dipped virtual density changes the nature of the autocrat's optimal policy. This makes it hard to compare the extent of information manipulation between the single-peaked and single-dipped cases.}

\section{Conclusion}

The growing literature on the rise of informational autocrats \citep{guriev/treisman:2022} discusses the modern autocrats' tendency to manipulate information. A natural question that follows from this research is about the conditions that make a society more susceptible to information manipulation. In this paper, we show that the dispersion of opinions puts a limit on an informational autocrat's ability to manipulate information.

To provide empirical support for this prediction, one needs to find variables that capture the dispersion of attitudes and opinions in a society. One readily available, albeit imperfect measure of heterogeneity is the Gini coefficient.\footnote{Although only a measure of income heterogeneity, the Gini coefficient has been shown to be related to social conflict \citep{rodrik:1999} and lack of social cohesion \citep*{easterly/woolcock/ritzen:2006}.} \cite{petrova:2008} provides evidence that suggests a link between income inequality and media freedom in autocracies. Figure 2 of \cite{petrova:2008} shows that, within autocracies (classified as countries with Democracy score $\leq 1$ in Polity IV dataset), there is a positive association between the Gini coefficient and Freedom House's media freedom index in 1994--2003. Reassuringly, the corresponding association is negative for countries classified as democracies in that period (Figure 1 of \citealp{petrova:2008}), suggesting that the lack of functioning democratic institutions is an essential part of this story.

Throughout our analysis, we considered the distribution of opinions and attitudes to be exogenous, and we remained agnostic about the forces that may increase its dispersion. Two channels that may lead to increased dispersion are independent media and online media. In a recent working paper, \citet*{enikolopov/rochlitz/schoors/zakharov:2023} demonstrate that access to independent online TV in Russia before the 2016 elections had asymmetric effects on individuals who rely on news from social media. Specifically, it bolstered the support among supporters of the regime while leading to a decline in support among those who opposed the regime. Motivated by their findings, and in light of the discussion here, one can argue that online media not only affect the attitudes of citizens but also have an impact on the effectiveness of traditional state-controlled media. In particular, online media do not have to convince every citizen---as long as they influence the opinions of {\it some} citizens, they could make it harder for the informational autocrat to engage in information manipulation.

In this paper, we focused on information manipulation as the only tool available to an autocrat. In reality, many autocrats have other tools at their disposal, such as repression and indoctrination \citep*{gitmez/sonin:2023,gehlbach/luo/shirikov/vorobyev:2022}, even if they do not always use them. The question of how the mix of tools used by autocrats is affected by the distribution of opinions is a fruitful avenue for future research.

\newpage

\appendix

\begin{huge}
\textbf{Appendices}
\end{huge}

\section{Proofs for Section \ref{section:monolithic}}

Because $f$ is continuously differentiable over its support and bounded, $h'(\mu)$ exists and is continuous. We begin by noting that the virtual density is single-peaked if and only if $h'(\mu)$ satisfies the {\it strict single-crossing-from-above property}. The strict single-crossing property is adapted from \citep[p.160]{milgrom/shannon:1994} and is as follows: 
\begin{center}
    If $h'(\mu) \geq 0$ for some $\mu \in [0,1]$, then $h'(\tilde{\mu}) > 0$ for all $\tilde{\mu} < \mu$.
\end{center}
In our proofs, we rely on the equivalence of this condition with single-peakedness of $h$.

\begin{proof}[Proof of Proposition \ref{proposition:single-peaked}]
If $h'(\mu)$ satisfies the strict single-crossing-from-above condition, by definition, so does $v''(\mu)$. Therefore, whenever $v(\mu)$ is convex at $\mu$, it is strictly convex at any $\tilde{\mu}< \mu$. This means that $v(\mu)$ is first strictly convex and then strictly concave. Therefore, the set where the concave closure of $v(\mu)$---call it $V(\mu)$---coincides with $v(\mu)$ has the following form:
\begin{align*}
    \{\mu\in [0,1]: V(\mu) = v(\mu)\} = \{0\} \cup [\hat{\mu},1],
\end{align*}
for some $\hat{\mu}\in[0,1]$. 

When $p_s < \hat{\mu}$, by Corollary 2 of \citet*{kamenica/gentzkow:2011}, the optimal policy generates two posteriors: $\mu \in \{0,\hat{\mu}\}$. This is achieved by two messages, with one perfectly revealing the bad state.

When $p_s \geq \hat{\mu}$, the optimal policy is not revealing any information. This can also be achieved by two messages, $m\in \{0,1\}$, and an information structure where $\Pr(m=1|\theta=0) = \Pr(m=1|\theta=1)$.  Message $m = 0$ will occur with probability zero, and the posterior beliefs following $m=0$ will be free in a Perfect Bayesian Equilibrium. One can assign the posterior $\Pr_r(\theta=0|m=0) = 1$ assign $m=0$ as the message that perfectly reveals the bad state.
\end{proof}

\begin{proof}[Proof of Proposition \ref{proposition:commonpriors-singlepeaked}]
Since $p_r=p_s$ for all $r$, equation \eqref{eqn:valuefunction} simplifies to
\[
v(\mu) = \int_0^{c(\mu,p_s)}f(c)dc.
\]
On the other hand, by definition, $c(\mu,p_s)=\mu$ for all $\mu$. Therefore, $h(\mu) = v'(\mu) = f(c(\mu,p_s))=f(\mu)$, and so, $h$ is single-peaked if and only if $f$ is single-peaked.
\end{proof}

\begin{proof}[Proof of Proposition \ref{proposition:commoncost-logconcave}]
When $c_r=c$ for all $r$, equation \eqref{eqn:valuefunction} simplifies to 
\begin{equation}\label{eqn:valuefunction2}
v(\mu) = \int_{p(\mu,c)}^1 f(p)dp,
\end{equation}
where
\begin{align} \label{eqn:pbound}
    p(\mu,c) \equiv \frac{1-\mu}{(1-\mu) + \mu \frac{1-c}{c}\frac{1-p_s}{p_s}}.
\end{align}
The virtual density $h(\mu)$ is then given by
\begin{align*}
    h(\mu) = v'(\mu)= - f(p(\mu,c)) \cdot\frac{\partial}{\partial \mu}p(\mu,c),
\end{align*}
and so,
\begin{align*}
    h'(\mu) = - f'(p(\mu,c)) \left(\frac{\partial}{\partial \mu}p(\mu,c)\right)^2 - f(p(\mu,c)) \cdot  \frac{\partial^2}{\partial \mu^2}p(\mu,c).
\end{align*}
Therefore, the sign of $h'(\mu)$ is the same as the sign of 
\[
-\frac{f'(p(\mu,c))}{f(p(\mu,c))}-\frac{\frac{\partial^2}{\partial \mu^2}p(\mu,c)}{\left(\frac{\partial}{\partial \mu}p(\mu,c)\right)^2},
\]
where $f(p(\mu,c))>0$ by assumption and $\partial p(\mu,c)/\partial \mu>0$ follows $\gamma>0$. Using \eqref{eqn:pbound} and substituting $\gamma = \frac{1-c}{c} \frac{1-p_s}{p_s}$, we get
\begin{align}
-\frac{f'(p(\mu,c))}{f(p(\mu,c))}-\frac{\frac{\partial^2}{\partial \mu^2}p(\mu,c)}{\left(\frac{\partial}{\partial \mu}p(\mu,c)\right)^2}& = -\frac{\partial}{\partial p} \log f(p(\mu,c))   -2\frac{\gamma-1}{\gamma}(1+(\gamma-1)\mu).\label{eq:sign_of_h_prime}
\end{align}
Substituting the value of $\gamma$ into equation \eqref{eqn:pbound} gives: $p(\mu,c) = \frac{1-\mu}{1+(\gamma-1)\mu}$. Solving for $\mu$,
\begin{align}
\mu = \frac{1-p(\mu,c)}{1+(\gamma-1)p(\mu,c)}.
\end{align}
Substituting for $\mu$ in equation \eqref{eq:sign_of_h_prime}, we get
\begin{align*}
-\frac{f'(p(\mu,c))}{f(p(\mu,c))}-\frac{\frac{\partial^2}{\partial \mu^2}p(\mu,c)}{\left(\frac{\partial}{\partial \mu}p(\mu,c)\right)^2} & = -\frac{\partial}{\partial p} \log f(p(\mu,c)) -2\frac{\gamma-1}{1+ (\gamma-1)p(\mu,c)}\\ & =  -\frac{\partial}{\partial p} \log f(p(\mu,c))+g\left(p(\mu,c)\right), 
\end{align*}
where $g(p) \equiv -2\frac{\gamma-1}{1+ (\gamma-1)p}$. Note that $g(p)$ is increasing in $p$, is convex in $p$ if $\gamma\leq 1$, and is concave in $p$ if $\gamma \geq 1$. Therefore, 
\begin{align} \label{eqn:gprime-min}
    \min_{p\in [0,1]} g'(p) & = \begin{cases}
    g'(0) & \text{ if }\;\gamma \leq 1,\\
    g'(1) & \text{ if }\; \gamma \geq 1
    \end{cases} = \begin{cases}
    2 (\gamma-1)^2 & \text{ if }\; \gamma \leq 1,\\
    2 \frac{(\gamma-1)^2}{\gamma^2} & \text{ if }\; \gamma \geq 1
    \end{cases} = 2 (\gamma-1)^2 \min\bigg\{1,\frac{1}{\gamma^2}\bigg\}.
\end{align}
If condition \eqref{eqn:commoncost-logconcave} holds, then
\[
\frac{\partial^2}{\partial p^2} \log f(p) < \min_{p\in [0,1]} g'(p),
\]
which implies
\begin{align} \label{eqn:logfflatter}
    \frac{\partial^2}{\partial p^2} \log f(p) < g'(p) \quad \forall p \in [0,1].
\end{align}
Our claim is that, under condition \eqref{eqn:commoncost-logconcave}, $h'(\mu)$ satisfies the strict single-crossing-from-above condition. To see this, take any two $\mu, \hat{\mu}$ with $\hat{\mu} < \mu$ and $h'(\mu) \geq 0$. Because $p(\mu,c)$ is strictly decreasing in $\mu$, $p(\mu,c) < p(\hat{\mu},c)$. Since $h'(\mu) \geq 0$, $\frac{\partial}{\partial p} \log f(p(\mu,c)) - g\left(p(\mu,c)\right) \leq 0$. Then,
\begin{align*}
    \frac{\partial}{\partial p}& \log f(p(\hat{\mu},c)) - g\left(p(\hat{\mu},c)\right)\\ 
    & = \frac{\partial}{\partial p} \log f(p(\mu,c)) - g\left(p(\mu,c)\right) + \int_{p(\mu,c)}^{p(\hat{\mu},c)} \underbrace{ \left( \frac{\partial^2}{\partial p^2} \log f(p) - g'\left(p\right) \right) }_{< 0 \text{ by } (\ref{eqn:logfflatter})} dp\\
    & < \frac{\partial}{\partial p} \log f(p(\mu,c)) - g\left(p(\mu,c)\right) \leq 0.
\end{align*}
Therefore, $h'(\hat{\mu}) > 0$. The result follows.
\end{proof}

We continue with some notation and preliminary results for the proof of Theorem \ref{theorem:sc-virtualdensitypolarization}.

\begin{lemma}\label{lemma:limits_of_v_prime}
The value function $v(\mu)$ satisfies
\[
\lim_{\mu\to 0}\mu v'(\mu) = \lim_{\mu\to 1}(1-\mu)v'(\mu)=0.
\]
\end{lemma}

\begin{proof}
First, note that
\begin{align*}
v'(\mu) & = \int_0^1 f(p,c(\mu,p)).\frac{\partial}{\partial \mu}c(\mu,p)dp \\ &   = \int_0^1 f(p,c(\mu,p))\frac{p(1-p)p_s(1-p_s)}{(p_s(1-p)+\mu(p-p_s))^2}dp,
\end{align*}
But since $f$ is bounded, there exist some $C>0$ such that
\begin{align*}
|v'(\mu)| & \leq C \int_0^1 \frac{p(1-p)p_s(1-p_s)}{(p_s(1-p)+\mu(p-p_s))^2}dp\\ & = C\frac{p_s(1 - p_s)}{(p_s - \mu)^3}\left[2(\mu-p_s) - \left(\mu(1-p_s) + p_s(1-\mu)\right)\log\left(\frac{\mu(1-p_s)}{p_s(1-\mu)}\right)\right].
\end{align*}
On the other hand,
\begin{align*}
& \lim_{\mu\to 0}\mu\cdot\frac{p_s(1 - p_s)}{(p_s - \mu)^3}\left[2(\mu-p_s) - \left(\mu(1-p_s) + p_s(1-\mu)\right)\log\left(\frac{\mu(1-p_s)}{p_s(1-\mu)}\right)\right]\\ & \qquad =\lim_{\mu\to 0}\frac{-(1 - p_s)}{p_s}\mu \log(\mu) = 0,
\end{align*}
and
\begin{align*}
& \lim_{\mu\to 1}(1-\mu)\cdot\frac{p_s(1 - p_s)}{(p_s - \mu)^3}\left[2(\mu-p_s) - \left(\mu(1-p_s) + p_s(1-\mu)\right)\log\left(\frac{\mu(1-p_s)}{p_s(1-\mu)}\right)\right]\\ & \qquad =\lim_{\mu\to 1}\frac{-p_s}{1-p_s}(1-\mu)\log(1-\mu) = 0.
\end{align*}
Therefore,
\[
\lim_{\mu\to0}\mu v'(\mu)=\lim_{\mu \to 1} (1-\mu)v'(\mu) = 0.
\]
This completes the proof of the Lemma.
\end{proof}

Consider a single-peaked virtual density $h(\mu)$. As discussed in the proof of Proposition \ref{proposition:single-peaked}, $\{\mu\in [0,1]: V(\mu) = v(\mu)\} = \{0\} \cup [\hat{\mu},1]$ for some $\hat{\mu} \in [0,1]$. Note that:
\begin{itemize}
     \item $v'(\mu)  \mu < v(\mu)$ for all $\mu \in (0,1)$ if and only if  $\hat{\mu} = 0$.
    \item $v'(\mu)  \mu > v(\mu)$ for all $\mu \in (0,1)$ if and only if  $\hat{\mu} = 1$.
    \item When $\hat{\mu} \in (0,1)$, it satisfies:
\begin{align} \label{eqn:muhat}
    v'(\hat{\mu}) \hat{\mu} = v(\hat{\mu}).
\end{align}
\end{itemize}
Let
\begin{align} \label{eqn:y}
    y(\mu) \equiv v'(\mu) \mu - v(\mu) = h(\mu) \mu - \int_{0}^{\mu}h(\tilde{\mu}) \tilde{\mu}, \qquad \forall \mu\in[0,1].
\end{align}
Then, $\hat{\mu} \in (0,1)$ is characterized by the equation: $y(\hat{\mu}) = 0$. We start with some remarks that will be used in the proof of Theorem \ref{theorem:sc-virtualdensitypolarization}.

\begin{remark} \label{remark:y0is0}
\textit{$\lim_{\mu\to 0}y(\mu) = 0$.} This follows from Lemma \ref{lemma:limits_of_v_prime} and the fact that $v(0) = 0$.
\end{remark}

\begin{remark} \label{remark:yiscontinuous}
\textit{$y(\mu)$ is continuous in $\mu$ over $(0,1)$.} This is because $f$ is continuous over its support.
\end{remark}

\begin{remark} \label{remark:yisbellshaped}
\textit{$y(\mu)$ is first strictly increasing and then strictly decreasing.} This is because $y'(\mu) = v''(\mu) \mu + v'(\mu) - v'(\mu) = v''(\mu) \mu = h'(\mu) \mu$. Since $h'(\mu)$ satisfies strict single crossing from above, so does $y'(\mu)$, and the remark follows.
\end{remark}

\begin{proof}[Proof of Theorem \ref{theorem:sc-virtualdensitypolarization}]
Take two single-peaked virtual densities $h_1(\mu)$ and $h_2(\mu)$ that satisfy equation \eqref{eqn:virtualdensity-polarization}. The cdf of $h_2$ is:
\begin{align*}
    H_2(\mu) \equiv \int_{0}^{\mu} h_2(x) dx = \int_{0}^{\mu} \alpha\left( h_1(x) \right) dx.
\end{align*}
For $k \in \{1,2\}$, let 
\begin{align*}
    y_k(\mu) \equiv h_k(\mu)\mu - H_k(\mu).
\end{align*}
Since both $h_1$ and $h_2$ are single-peaked, Remarks \ref{remark:y0is0}, \ref{remark:yiscontinuous} and \ref{remark:yisbellshaped} imply that the set $\mathcal{U}_{y_k}\equiv \{\mu\in[0,1]: y_k(\mu)\geq 0\}$, for $k\in \{1,2\}$, has the following form:
\begin{align*}
    \mathcal{U}_{y_k}=[0,\hat{\mu}_k].
\end{align*}
The proof goes through showing that $\hat{\mu}_2 \leq \hat{\mu}_1$. If $\hat\mu_1 = 1$, this inequality is satisfied. If $\hat\mu_1 < 1$, $y_1(\hat\mu_1) = 0$, which implies: $\frac{H_1(\hat\mu_1)}{\hat\mu_1} = h_1(\hat\mu_1)$. Then,
\begin{align*}
    \frac{H_2(\hat\mu_1)}{\hat\mu_1} & = \frac{\int_{0}^{\hat\mu_1} \alpha\left( h_1(x) \right) dx}{\hat\mu_1} \geq \alpha\left(\frac{\int_{0}^{\hat\mu_1} h_1(x) dx}{\hat\mu_1}\right) = \alpha\left(\frac{H_1(\hat\mu_1)}{\hat\mu_1}\right) = \alpha\left( h_1(\hat\mu_1) \right) = h_2(\hat\mu_1),
\end{align*}
where the inequality follows from the integral form of Jensen's inequality (e.g., \citealt{dragomir/khan/addiselam:2016}). Therefore, $H_2(\hat\mu_1) \geq h_1(\hat\mu_1) \hat\mu_1$ and $y_2(\hat\mu_1) \leq 0$.

Since $\mathcal{U}_{y_2} = [0,\hat\mu_2]$, $y_2(\mu)$ crosses 0 from above at $\hat\mu_2$, and $y'_2(\hat\mu_2) \leq 0$. Since $y'_2(\mu)$ satisfies strict single-crossing from above, $y_2(\tilde\mu) < 0$ for any $\tilde\mu > \hat\mu_1$. We conclude that $\hat\mu_2 \leq \hat\mu_1$.

To conclude the proof, consider three cases:
\begin{enumerate}
    \item If $p_s \geq \hat{\mu}_1$, the optimal policy does not reveal any information in either case, and we pick $\sigma_1^0 = \sigma_2^0 = 1$.
    \item If $\hat{\mu}_1 > p_s \geq \hat{\mu}_2$, the optimal policy under $h_2(\mu)$ does not reveal any information. In this case, we pick $\sigma_2^0 = 1$ and $\sigma_1^0 < 1$.
    \item If $p_s < \hat{\mu}_2$, the optimal policies $\sigma_1^0$ and $\sigma_2^0$ satisfy:
    \begin{align*}
    \frac{p_s}{p_s + (1-p_s)\sigma_1^0} = \hat{\mu}_1, \qquad  \frac{p_s}{p_s + (1-p_s)\sigma_2^0} = \hat{\mu}_2.
\end{align*}
Then, $\hat{\mu}_1 \geq \hat{\mu}_2$ implies $\sigma_1^0\leq \sigma_2^0$.
\end{enumerate}
In any case, $\sigma_1^0\leq \sigma_2^0$, and the result follows.
\end{proof}

\section{Proofs for Section \ref{section:single-dipped}}

Note that single-dippedness of the virtual density is equivalent to the following {\it strict single-crossing-from-below property} for $h'(\mu)$:
\begin{center}
    If $h'(\mu) \geq 0$ for some $\mu \in [0,1]$, then $h'(\tilde{\mu}) > 0$ for all $\tilde{\mu}> \mu$.
\end{center}

\begin{proof}[Proof of Proposition \ref{proposition:single-dipped}]
If $h'(\mu)$ satisfies the strict single-crossing-from-below condition, by definition, so does $v''(\mu)$. Therefore, whenever $v(\mu)$ is convex at $\mu$, it is strictly convex at any $\hat{\mu}\geq \mu$. This means that $v(\mu)$ is first strictly concave and then strictly convex. Therefore, the set where the concave closure of $v(\mu)$ coincides with $v(\mu)$ has the following form:
\begin{align*}
    \{\mu\in [0,1]: V(\mu) = v(\mu)\} = [0,\hat{\mu}] \cup \{1\}.
\end{align*}
When $p_s < \hat{\mu}$, the optimal policy is not revealing any information. This can be achieved by two messages, $m\in \{0,1\}$, and an information structure where $\Pr(m=1|\theta=0) = \Pr(m=1|\theta=1)=0$. Message $m = 1$ will occur with probability zero, and the posterior beliefs following $m=1$ will be free in a Perfect Bayesian Equilibrium. One can assign posteriors $\Pr_r(\theta=1|m=1)= 1$ to make $m=1$ as the message that perfectly reveals the good state.

When $p_s \geq \hat{\mu}$, by Corollary 2 of \citet*{kamenica/gentzkow:2011}, the optimal policy generates two posteriors: $\mu \in \{\hat{\mu},1\}$. This is achieved by two messages, $m\in \{0,1\}$, where message $m=1$ perfectly reveals the good state.
\end{proof}

\begin{proof}[Proof of Proposition \ref{proposition:commonpriors-singledipped}]
The proof of Proposition \ref{proposition:commonpriors-singledipped} is identical to the proof of Proposition \ref{proposition:commonpriors-singlepeaked}.
\end{proof}

\begin{proof}[Proof of Proposition \ref{prop:commoncost-logconvex}]
The proof follows identical steps to that of Proposition \ref{proposition:commoncost-logconcave} until equation \eqref{eqn:gprime-min}. The rest of the argument is provided below.

Recall that $g(p) \equiv -2\frac{\gamma-1}{1+ (\gamma-1)p}$ and $g(p)$ is increasing in $p$, convex in $p$ if $\gamma\leq 1$, and concave in $p$ if $\gamma \geq 1$. Therefore, 
\begin{align} \label{eqn:gprime-max}
    \max_{p\in [0,1]} g'(p) = \begin{cases}
    g'(1) & \text{ if } \gamma \leq 1\\
    g'(0) & \text{ if } \gamma \geq 1
    \end{cases} = \begin{cases}
    2 \frac{(\gamma-1)^2}{\gamma^2} & \text{ if } \gamma \leq 1\\
    2 (\gamma-1)^2 & \text{ if } \gamma \geq 1
    \end{cases} = 2 (\gamma-1)^2 \max\bigg\{1,\frac{1}{\gamma^2}\bigg\}.
\end{align}
If condition \eqref{eqn:commoncost-logconvex} holds,
\[
    \frac{\partial^2}{\partial p^2} \log f(p) > \max_{p\in [0,1]} g'(p),
\]
and so
\begin{align} \label{eqn:logfsteeper}
    \frac{\partial^2}{\partial p^2} \log f(p) > g'(p) \quad \forall p \in [0,1].
\end{align}
Our claim is that, under condition \eqref{eqn:commoncost-logconvex}, $h'(\mu)$ satisfies the strict single-crossing-from-below condition. To see this, take any two $\mu, \hat{\mu}$ with $\hat{\mu} >\mu$ and $h'(\mu) \geq 0$. Because $p(\mu,c)$ is strictly decreasing in $\mu$, $p(\hat{\mu},c) < p(\mu,c)$. Since $h'(\mu) \geq 0$, $\frac{\partial}{\partial p} \log f(p(\mu,c)) - g\left(p(\mu,c)\right) \leq 0$. Then,
\begin{align*}
    \frac{\partial}{\partial p}& \log f(p(\hat{\mu},c)) - g\left(p(\hat{\mu},c)\right)\\ 
    & = \frac{\partial}{\partial p} \log f(p(\mu,c)) - g\left(p(\mu,c)\right) - \int_{p(\hat\mu,c)}^{p(\mu,c)} \underbrace{ \left( \frac{\partial^2}{\partial p^2} \log f(p) - g'\left(p\right) \right) }_{> 0 \text{ by } (\ref{eqn:logfsteeper})} dp\\
    & < \frac{\partial}{\partial p} \log f(p(\mu,c^*)) - g\left(p(\mu,c)\right) \leq 0.
\end{align*}
Therefore, $h'(\hat{\mu})>0$. The result follows.
\end{proof}

We continue by introducing some notation and preliminary results for the remaining proofs.

Consider a single-dipped virtual density $h(\mu)$. As discussed in the proof of Proposition \ref{proposition:single-dipped}, $\{\mu\in [0,1]: V(\mu) = v(\mu)\} = [0,\hat{\mu}] \cup \{1\}$ for some $\hat{\mu} \in [0,1]$. Note that:
\begin{itemize}
     \item $v'(\mu)  (1-\mu) > 1-v(\mu)$ for all $\mu \in (0,1)$ if and only if  $\hat{\mu} = 1$.
    \item $v'(\mu) (1-\mu) < 1-v(\mu)$ for all $\mu \in (0,1)$ if and only if  $\hat{\mu} = 0$.
    \item When $\hat{\mu} \in (0,1)$, it satisfies:
\begin{align} \label{eqn:muhat_SD}
    v'(\hat{\mu}) (1-\hat{\mu}) = 1-v(\hat{\mu}).
\end{align}
\end{itemize}
Let
\begin{align} \label{eqn:y_SD}
    z(\mu) \equiv v'(\mu)(1-\mu) - (1-v(\mu)) = h(\mu) (1-\mu) - \int_{\mu}^{1}h(\tilde{\mu}) \tilde{\mu}, \qquad \forall \mu\in[0,1].
\end{align}
Then, $\hat{\mu} \in (0,1)$ is characterized by the equation: $z(\hat{\mu}) = 0$. We start with some remarks.

\begin{remark} \label{remark:y0is0_SD}
\textit{$\lim_{\mu\to 1}z(\mu) = 0$.} This follows Lemma \ref{lemma:limits_of_v_prime} and the fact that $1-v(1) = 0$.
\end{remark}

\begin{remark} \label{remark:yiscontinuous_SD}
\textit{$z(\mu)$ is continuous in $\mu$ over $(0,1)$.} This is because $f$ is continuous over its support.
\end{remark}

\begin{remark} \label{remark:yisbellshaped_SD}
\textit{$z(\mu)$ is first strictly decreasing and then increasing.} This is because $z'(\mu) = v''(\mu) (1-\mu)-v'(\mu) + v'(\mu) = v''(\mu)(1-\mu) = h'(\mu)(1-\mu)$. Since $h'(\mu)$ satisfies strict single crossing from below, so does $z'(\mu)$, and the remark follows.
\end{remark}

\begin{proof}[Proof of Theorem \ref{theorem:sd-virtualdensitypolarization}]
Take two single-peaked virtual densities $h_1(\mu)$ and $h_2(\mu)$ that satisfy equation \eqref{eqn:virtualdensity-polarization_SD}. For $k \in \{1,2\}$, let 
\begin{align*}
    z_k(\mu) \equiv h_k(\mu)(1-\mu) - (1-H_k(\mu)).
\end{align*}
Based on Remarks \ref{remark:y0is0_SD}, \ref{remark:yiscontinuous_SD} and \ref{remark:yisbellshaped_SD}, the set $\mathcal{L}_{z_k}\equiv \{\mu\in[0,1]: z_k(\mu)\leq 0\}$ has the following form:
\begin{align*}
    \mathcal{L}_{z_k}=[\hat{\mu}_k,1].
\end{align*}
The proof goes through showing that $\hat{\mu}_2 \leq \hat\mu_1$. If $\hat\mu_1 = 1$, this inequality is satisfied. If $\hat\mu_1 < 1$, $z_1(\hat\mu_1) \leq 0$, which implies: $\frac{1-H_1(\mu))}{1-\hat\mu_1} \geq h_1(\hat\mu_1)$. Then,
\begin{align*}
    \frac{1-H_2(\hat\mu_1)}{1-\hat\mu_1} & = \frac{\int_{\hat\mu_1}^{1} \alpha\left( h_1(x) \right) dx}{1-\hat\mu_1} \geq \alpha\left(\frac{\int_{\hat\mu_1}^{1} h_1(x) dx}{1-\hat\mu_1}\right) = \alpha\left(1-\frac{H_1(\hat\mu_1)}{1-\hat\mu_1}\right) \geq \alpha\left( h_1(\hat\mu_1) \right) = h_2(\hat\mu_1).
\end{align*}
Therefore, $z_2(\hat\mu_1) \leq 0$. This means $\hat\mu_1 \in \mathcal{L}_{z_2} = [\hat\mu_2,1]$, and therefore $\hat\mu_2 \leq \hat\mu_1$. Repeating the same argument in the proof of Theorem \ref{theorem:sc-virtualdensitypolarization}, we conclude that $\sigma_2^1 \geq \sigma^1_1$.
\end{proof}

\newpage

\fontsize{10.1}{10.1}\selectfont\baselineskip0.385cm
\bibliographystyle{econ-aer}
\bibliography{bib-media}

\end{document}